\documentstyle[epsf,rotate]{mn}

\def\sec{{^\prime}{^\prime}} 
\def\min{^\prime}
\def\deg{^\circ}
\def\rasec{\hbox{$\,$\raise 0.6 ex \hbox{s}\kern-.35em
                  \lower 0.0 ex \hbox{.}$\,$}}        
\def\decsec{\hbox{$\,$\raise 0.0 ex \hbox{$\sec$}\kern-.45em
                  \lower 0.0 ex \hbox{.}$\,$}}         
\def\decmin{\hbox{$\,$\raise 0.0 ex \hbox{$\min$}\kern-.45em
                  \lower 0.0 ex \hbox{.}$\,$}}        
\def\gtabouteq{\,\hbox{\raise 0.5 ex \hbox{$>$}\kern-.77em 
                    \lower 0.5 ex \hbox{$\sim$}$\,$}}       
\def\ltabouteq{\,\hbox{\raise 0.5 ex \hbox{$<$}\kern-.77em 
                     \lower 0.5 ex \hbox{$\sim$}$\,$}}

\def\twto{$^{12}$CO(2--1) }
\def\thto{$^{13}$CO(2--1) }
\def\twtt{$^{12}$CO(3--2) }
\def\twoz{$^{12}$CO(1--0) }
\def\thco{$^{13}$CO }
\def\twco{$^{12}$CO }

\title{Molecular Gas in the Perseus Cooling Flow Galaxy, NGC~1275}

\author[Terry J. Bridges and Judith A. Irwin]
{Terry J. Bridges$^{1,2,4}$
and Judith A. Irwin$^{3,5}$ \\
$^1$Royal Greenwich Observatory,
Madingley Road,
Cambridge, England,
CB3 0EZ\\
$^2$Institute of Astronomy, Madingley Road, Cambridge, England, CB3 0HA \\
$^3$Department of Physics, Queen's University, Kingston, Ontario, Canada, K7L 3N6\\
$^4$E-mail:~~tjb@ast.cam.ac.uk \\
$^5$E-mail:~~irwin@astro.queensu.ca}

\date{Accepted~~~~~~~.  Received~~~~~~~~~}

\pubyear{1998}

\begin{document} 

\maketitle
\begin{abstract}

The central arcminute of the Perseus cooling flow galaxy, NGC~1275,
has been mapped with the JCMT in \twto at 21$\sec$ resolution,
with detections out to at least 36$\sec$ (12 kpc).  Within the
limits of the resolution and coverage, the distribution of gas 
appears to be roughly E-W, consistent with previous observations
of CO, X-ray, H$\alpha$, and dust emission.  The total detected 
molecular hydrogen mass is $\sim$ 1.6 $\times$ 10$^{10}$ M$_\odot$,
using a Galactic conversion factor.
The inner central rotating disk
is apparent in the data, but the overall distribution is not one
of rotation.  Rather, the line profiles are bluewards asymmetric,
consistent with previous observations in HI and [OIII].  We suggest
that the blueshift may be due to an acquired mean velocity
of $\sim$ 150 km s$^{-1}$ imparted by the radio jet 
in the advancing direction.  Within the uncertainties
of the analysis, the available radio energy appears to be sufficient,
and the interpretation is consistent with that  of
Bohringer et al. (1993) for displaced X-ray emission.  
 
We have also made the first observations of
\thto and \twtt emission from the central 21$\sec$
region of NGC~1275 and combined these data with
IRAM data supplied by Reuter et al. (1993) to form line ratios
over equivalent, well-sampled regions.  An LVG radiative
transfer analysis indicates that the line ratios are not well
reproduced by a single value of kinetic temperature, molecular
hydrogen density, and abundance per unit velocity gradient.
At least two temperatures are suggested by a simple two-component
LVG model, possibly reflecting a temperature gradient in this
region.

\end{abstract}


\section{Introduction}

Cluster cooling flows present a most intriguing conspiracy.  X-ray
observations have revealed large quantities of hot, diffuse gas in
most if not all clusters (e.g. Fabian 1994).   In many clusters
(70--90\%) the central gas densities are high enough that the radiative
cooling
time becomes shorter than the Hubble time (e.g. Edge, Stewart, \&
Fabian 1992; Fabian 1994).  Thus pressure-driven cooling flows are
inferred.  In several clusters, soft X-ray emission has also
revealed the presence of
gas at intermediate temperatures of 10$^5$--10$^7$ K, as would be
expected for gas condensing out of a cooling flow (e.g. Johnstone
et al. 1992).  If the inferred mass deposition rates of
between 10--1000 M$_\odot$ are maintained  over a Hubble
time, then $\sim$ 10$^{12}$ -- 10$^{13}$ M$_\odot$ of material should
be deposited in the cores of clusters.  

However, there is little evidence for all of this supposed
infalling material at other wavelengths.  Optical data apparently
rule out star
formation with a `normal' (ie. Galactic) mass function
(e.g. McNamara \& O'Connell 1992).  Some fraction of the
cooling material should exist in cool atomic and/or molecular
clouds.
However, HI emission has not been detected in any cooling flow
cluster, with upper limits of 10$^9$--10$^{10}$ M$_\odot$, while
HI has been detected in absorption in only four clusters (e.g.
Jaffe 1992).
As well, Annis \& Jewitt (1993) carried out an unsuccessful search
for thermal emission from cold dust in 11 clusters, placing upper
limits of $\sim$ 10$^{10}$ M$_\odot$ on the mass of cold gas in
these clusters, assuming a Galactic gas/dust ratio.  Searches for
molecular gas (ie. CO) have also been unfruitful, with one
exception (see below)
in over a dozen clusters (e.g. Grabelski \& Ulmer 1990; McNamara \&
Jaffe 1994); again typical upper limits to the H$_2$ mass are
$\sim$ 10$^{10}$ M$_\odot$.

The only cooling flow galaxy in which CO has yet been detected
is the one in the Perseus Cluster (Abell 426) for which infall rates of
$\sim$ 300--500 M$_\odot$ per year centered on the giant cD galaxy, 
NGC~1275, have been inferred
 (Gorenstein et al. 1978; Fabian et al. 1981; Mushotzky
et al. 1981; Allen et al. 1992; Allen \& Fabian 1997).  
Allen \& Fabian (1997) find excess X-ray absorption 
from ROSAT data in the central 30\arcsec~ of the cluster implying 
a total mass of $\sim$ 4 $\times$ 10$^8$ M$_\odot$.
A system of H$\alpha$ filaments extending to $\approx$ 2$\min$ = 40
kpc from the core of NGC~1275 has been linked to the cooling gas
(Lynds 1970; Cowie, Fabian, \& Nulsen 1980; Heckman et al. 1989).  
NGC 1275 is also home to the highly variable
radio source 3C84 (e.g. Pedlar et al. 1990). HI, in quantities of
$\sim$ 10$^{10}$ M$_\odot$, has been detected in absorption against
this source (Jaffe 1990) and the relativistic outflow appears
to be displacing X-ray emitting gas on arcminute scales
(Bohringer et al. 1993).  A foreground infalling 
($\Delta$ V $\approx$ 3000 km s$^{-1}$) system is also known to
exist (Minkowski 1957;
Lynds 1970; Hu et al. 1983; Caulet et al. 1992) and there is
some evidence that NGC~1275 has experienced a previous merger
(e.g. Holtzman et al. 1992).
Braine et al. (1995) attribute the origin of the
molecular gas in NGC~1275 to a previous merger as do Lester et al.
(1995) for the origin of the 100 $\mu$m emitting dust.

Previous CO observations of NGC~1275
 include single-dish detections by
Lazareff et al. (1989), and Mirabel, Sanders, \& Kazes (1989), as well as a
partial mapping by Reuter et al. (1993) to $\sim$ 20$^{\prime
\prime}$ radius in $^{12}$CO(1--0)
and $^{12}$CO(2--1) with the Institut
de Radioastronomie Millim\'etrique (IRAM) telescope.  Estimated
H$_2$ masses are of order M$_{H_2}$ $\sim$ 10$^{10}$ M$_\odot$
using Galactic CO to H$_2$ conversion factors.
High resolution 
$^{12}$CO(1--0) interferometry has now also been carried out
by Braine et al. (1995) and Inoue et al. (1996).  The latter
authors find
a rotating ring-like CO structure with a ring radius of 
3.5$^{\prime \prime}$.
A hot (T = 1700 K) component of molecular hydrogen has also been
directly detected through 2 $\mu$m emission within the central few
arcseconds (Fischer et al. 1987; Kawara \& Taniguchi 1993; Inoue et
al. 1996).  No attempt has yet been made to determine the physical
properties of the molecular gas in this most unusual system.

In this paper, we present $^{12}$CO(2--1), $^{13}$CO(2--1), and
$^{12}$CO(3--2) observations of NGC 1275 using the 15 m James Clerk
Maxwell Telescope (JCMT).  We have mapped the galaxy in the 
$^{12}$CO(2--1) line out to $\sim$ 1 arcmin and in the
$^{12}$CO(3--2) line out to $\sim$ 14 arcsec.  In addition, we have obtained 
IRAM $^{12}$CO(2--1) and $^{12}$CO(1--0) data, kindly supplied by
Reuter et al. (1993).  Together with the JCMT data, we perform a
line ratio analysis for the central position in order to constrain
the physical conditions in the molecular gas and see whether these
are consistent with a cooling flow origin.  
The observations are
described in Section 2, the results are presented in Section 3, the
line ratio analysis is discussed in Section 4, some physical quantities
are derived in Section 5, there is a Discussion in Section 6, and
our main conclusions are summarized in Section 7.
For NGC~1275, we take D = 70 Mpc,
(cz = 5250 km s$^{-1}$; H$_0$ = 75 km s$^{-1}$ Mpc$^{-1}$) in which
case 1$\sec$ = 340 pc.

\section{Observations and Data Reduction}

\twto and \thto data were obtained during 3 shifts at the 
JCMT{\footnote{The James Clerk Maxwell Telescope is operated
by the Royal Observatory, Edinburgh on behalf of the Science
and Engineering Research Council of the United Kingdom, the
Netherlands Organization for Scientific Research, and the
National Research Council of Canada.}}
on Mauna Kea during 23--25 November 1994.  The beam-size at
230 GHz (\twto) is 21\arcsec and is 22\arcsec at 220 GHz (\thto),
which corresponds to $\sim$ 7 kpc for H$_0$ = 75 km s$^{-1}$ 
Mpc$^{-1}$.
From scans of Saturn, 
the main-beam efficiency, $\eta_{mb}$, for \twto and \thto were
determined to be 0.69 $\pm$ 0.05 and 0.66 $\pm$ 0.05 respectively;
these values agree well with that determined by
JCMT staff between 1 Nov 1994 to Jan 30 1995 (0.69 $\pm$ 0.01; see
JCMT WWW page).  Unless otherwise indicated, all
temperatures quoted below are on the
main-beam brightness temperature scale T$_{mb}$.  The system 
temperature T$_{sys}$ (in T$_a^*$ units)
ranged between 500--600K for \twto and
600--700K for \thto. 
Pointing was done on the central radio
source 3C84, CRL2688 and Saturn several times per night. 
During
an 8 hour run, we found rms drifts in pointing of 3-4 arcsec 
on average.
We chose our
(0,0) position to be the same as that of Reuter et al. (1993), i.e.
$\alpha$= 03 16 29.29, $\delta$=$+$41 19 51.9 (1950), in order to
facilitate the line ratio analysis.  This position is 3.5\arcsec
west of the galaxy center which is at the 3C84 radio core,
i.e. RA(1950) =
3$^h$ 16$^m$ 29$\rasec$6, DEC(1950) =41$\deg$ 19$\min$ 52$\sec$
(Pedlar et al. 1990).  The displacement is $<<$ the beam size.

Integrations were carried out using receiver RxA2 with the DAS in
its 760 
Mhz configuration ($\sim$ 1000 km s$^{-1}$, 0.8 km s$^{-1}$ per channel).
We beam-switched with a 1 hz chopping frequency and a 
150\arcsec throw. 
CRL 2688 scans were used to check the calibration
for \twto and \thto. 
Two \twto scans of CRL 2688 yield integrated intensities of 
232 K km s$^{-1}$ (Nov 23, T$_a^*$ units) and 202 K km s$^{-1}$ (Nov 25),
as compared to the published value of 168 K km s$^{-1}$ on October 20, 1994.
For \thto the measured 
and published line strengths are 42 and 46 K km s$^{-1}$ respectively, in
excellent agreement.  Throughout later observing runs, we also
continued to check that the (0,0) spectrum gave consistent
results.
We integrated for 5040  
sec at the galaxy center in \twto (Fig. 1), and 5520 sec
in \thto (Fig. 3).  

Subsequent data reduction was done using
the UNIX SPECX package developed by Rachael Padman.  The spectra were averaged
(weighted inversely by T$_{sys}$),
20-channel binned (giving a final velocity resolution of 16.25
km s$^{-1}$),
and baseline subtracted (linear baselines were mostly
used); for \twto only 3720 sec of
data were usable.  We also carried out a partial mapping in \twto,
with integrations ranging between 1800--3600 sec for 8
points out to 1 arcmin from the galaxy center (see below and
Section 3.3 for more details).  
The sampling was 20$\arcsec$
in RA and 15$\arcsec$ in DEC with beam centers in each row
displaced by 10$\arcsec$ so that full coverage was obtained
with the 21$\arcsec$ beam, though not at the Nyquist rate.

\twtt observations of a 5-point (0,0; $-$7,$+$7; $+$7,$-$7;
$+$7,$+$7; $-$7,$-$7) map at the center of NGC 1275 were obtained
during service time on July 31 1995.  The beamsize is 
14$^{\prime \prime}$ at 345 Ghz, and $\eta_{mb}$= 0.525 $\pm$
0.033, as measured from Saturn scans.
Receiver RxB3i was used with the DAS in its 760 Mhz configuration
($\sim$ 600 km s$^{-1}$, 0.54 km s$^{-1}$ per channel).  A total of 3000 sec
at 0,0 and 2400 sec at each of the other four points was obtained.
Pointing done on W3, G34.3, CRL618, and Saturn shows $\sim$ 3\arcsec
drifts over the 8 hour shift.
Again we have averaged, binned by 20 channels (giving a velocity
resolution
of 11 km s$^{-1}$), and baseline-subtracted.  

Finally, during 7 shifts in December 1995, we extended our \twto
mapping by obtaining 12 further points extending out to $\sim$
1 arcmin from the galaxy center; integrations were typically 5400
sec (see Section 3.3).  Receiver A2 was used
in 750 Mhz configuration 
(0.8 km s$^{-1}$ per channel), with
1 hz beamswitching and a 150\arcsec throw.  Spectra were averaged and
20-channel binned for a final velocity resolution of 16.25 km s$^{-1}$.
Saturn scans were again used to determine 
a $\eta_{mb}$ value of 0.61 $\pm$ 0.05 for \twto on Dec 4/5, where
the uncertainty includes known random errors.  
The JCMT value is given as 0.69 $\pm$ 0.01 (Jan 25/96), but the
uncertainty in a single measurement is closer to $\pm$ 0.07
(Per Friberg, private comm.) 
where changes could be due to a variety of sources, including 
sky variations, standing waves, uncertainties in beamsize, and
errors in pointing and focussing (Henry Matthews, private comm.). 
Thus it is important to measure $\eta_{mb}$ during each observing
session, as we have done.  We found the same value of 
$\eta_{mb}$ for each of
three Saturn scans over Dec 4/5.
T$_{sys}$ for \twto
ranged between 300--400 K (Dec 3) to 500--600 K (Dec 4/5).  Pointing
done on W3, 3C84 and Saturn show $\sim$ 3\arcsec rms drifts over
an 8-hour shift.  \twto calibration was checked using scans of NGC
1275, which also allowed us to check consistency with our previous 
data taken in Dec 1994.   Five 600 sec scans taken at 0,0 yield
integrated line strengths agreeing with each other to within 25\%,
and also with our December 1994 data (e.g. Figure 1) to within 25\%.


The IRAM data were supplied by H.-P. Reuter (private communication) and
full details can be found in Reuter et al. (1993). 

A summary of the spectral lines and telescope parameters is listed in Table 1.

\begin{table}
 \caption{Telescope Parameters}
 \label{symbols}
 \begin{tabular}{cccl}
  Spectral Line  &  Telescope & Beam FWHM  & $\eta_{mb}$ \\
 $^{12}$CO(1-0)  &  IRAM & 21$\arcsec$ &  0.67 \\
 $^{12}$CO(2-1)  &  IRAM & 12$\arcsec$ &  0.5 \\
 $^{12}$CO(2-1)  &  JCMT & 21$\arcsec$ &  0.69/0.61$^1$ \\
 $^{13}$CO(2-1)  &  JCMT & 22$\arcsec$ &  0.66 \\
 $^{12}$CO(3-2)  &  JCMT & 14$\arcsec$ &  0.53 \\
 \end{tabular}
 \bigskip
\noindent$^1$~~For December 1994 and December 1995 respectively

\end{table}

\section{Results}

\subsection{Spectra}

The \twto and \twtt spectra
are shown in Fig. 1 and 2, respectively. 
For point (+20,0) of Fig. 1, note that
the low velocity end of the baseline has been truncated.  This is 
because several of the individual scans at this position
were centered at an offset velocity,
necessitating a shift in the velocity coordinate before averaging.  The
non-overlapping baseline regions were 
then truncated, but the integrity of the
spectral line itself (based on inspection of all individual scans
with full baselines) has
been retained. 
The \thto spectrum for
the center position is shown in Fig. 3 with the central \twto
spectrum superimposed.  We consider the isotopic line to be a marginal
detection (cf. the similarly placed peaks and troughs in comparison to
\twto), but given the low S/N, we place little weight upon
this particular spectrum in subsequent analysis. 

The strong radio source, 3C84, occurs at the (0,0) position (Pedlar
et al. 1990) which has a core strength in the range 2 - 6 Jy (Braine et al.
1995; private communication to Lester 1995; Inoue et al. 1996) at 115 GHz
and a total size less than 10 mas (Readhead et al. 1983), or area filling
factor $<$ 2 $\times$ 10$^{-7}$ in our 21$\sec$ \twto beam.
An inspection of the original spectra before and after binning/average
shows no compelling evidence for 
absorption features above the noise level. This is confirmed by
Braine et al. (1995) and Inoue et al. (1996), neither of whom
detect any \twoz absorption in much smaller interferometric
beams of 2$\sec$ and 4$\sec$, respectively.

\begin{figure*}
\epsfysize 6.0 truein
\hfil\rotate[r]{\epsffile
{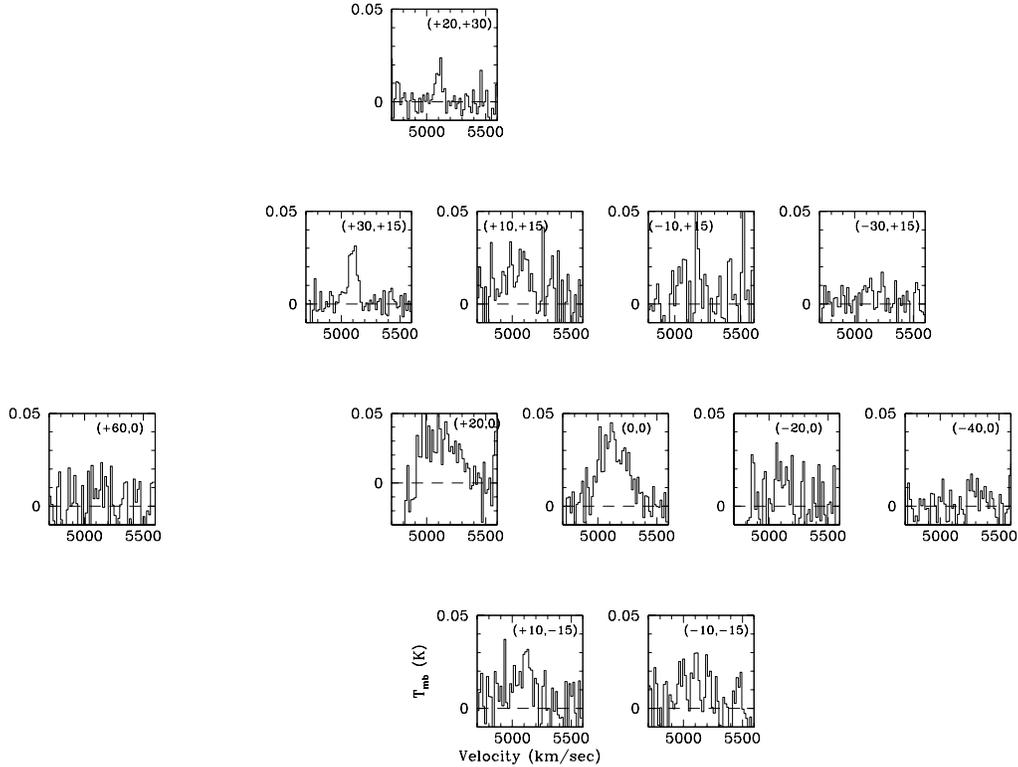}}\hfil
\caption{\twto spectra at the various observing positions.
The temperature scale is in T$_{mb}$.  
We show only detections and suspected detections.}
\label{fig1}
\end{figure*}


\begin{figure*}
\epsfysize 6.0 truein
\hfil{\epsffile{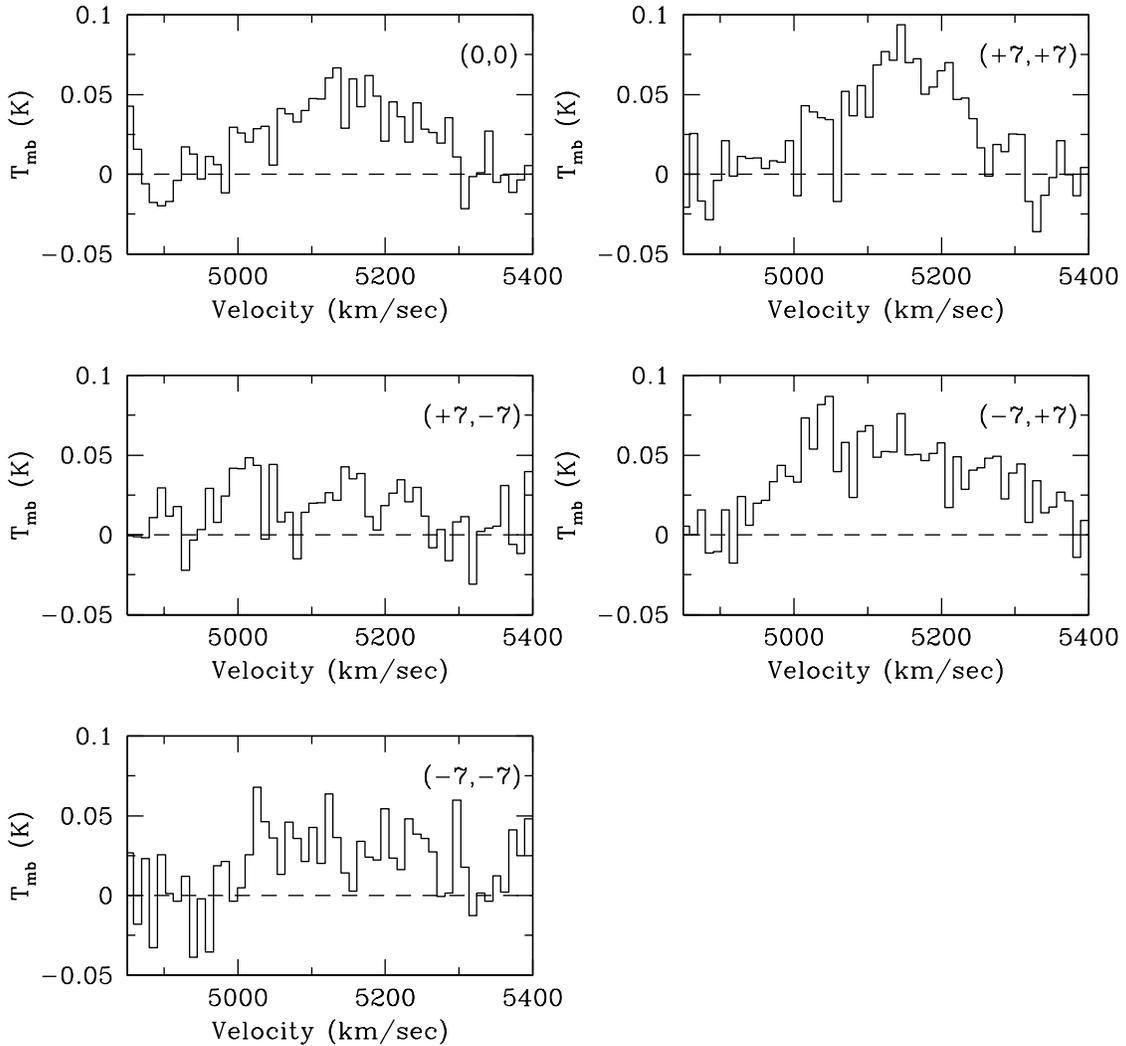}}\hfil
\caption{ \twtt spectra immediately around the center}
\label{fig2}
\end{figure*}

\begin{figure*}
\epsfysize 4 truein
\hfil{\epsffile{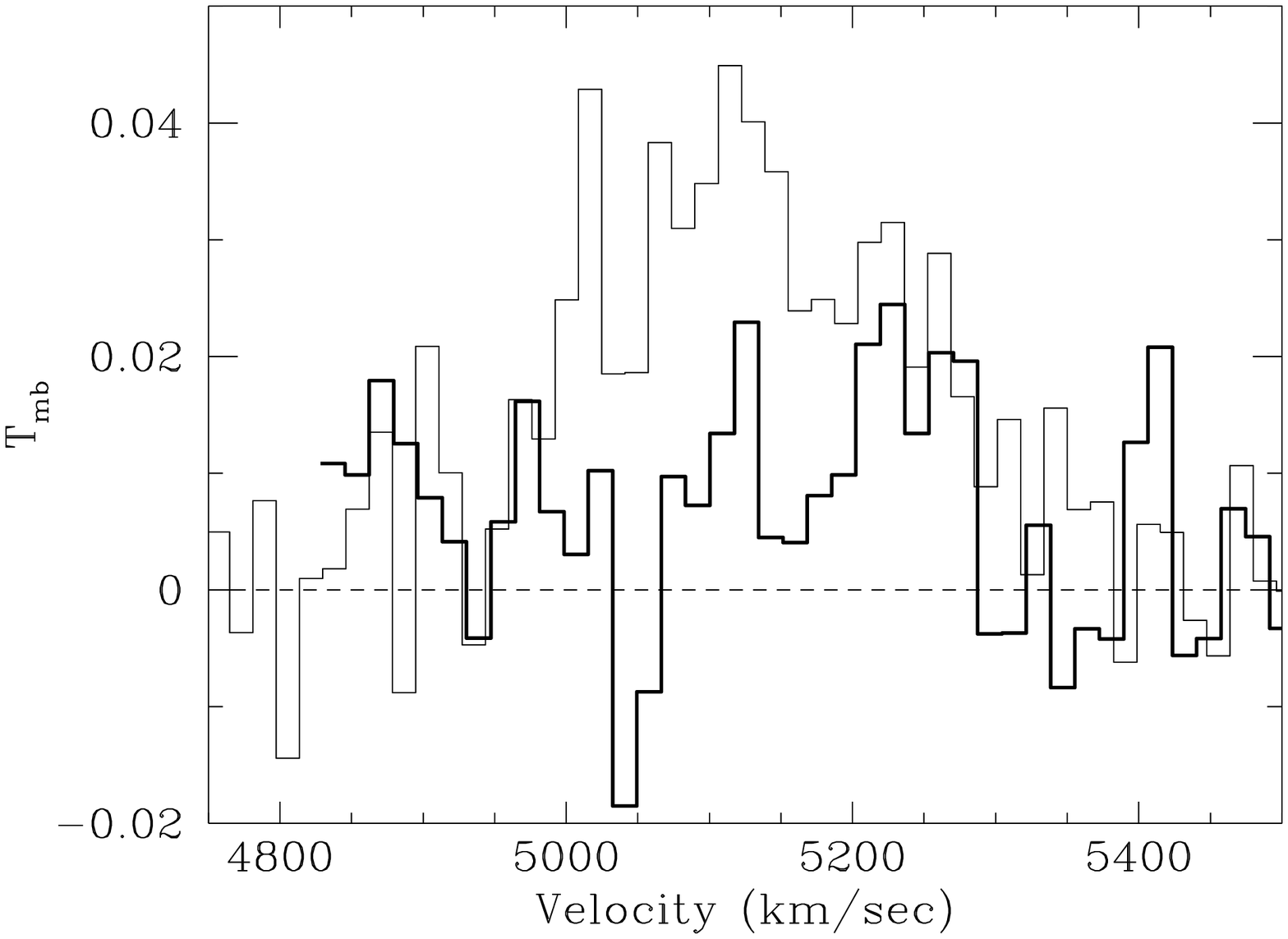}}\hfil
\caption{\thto spectrum at the (0,0) position (heavy line) with the
\twto spectrum (light line) superimposed.  The apparent increase in emission
at the low velocity end of the spectrum is due to baseline curvature.}
\label{fig3}
\end{figure*}

The integrated line intensities at the
measured positions, $\int T_{mb}\,dV$, are given in Table 2, along
with the peak T$_{mb}$ and FWHM.  The quoted uncertainties take
into account the noise in the spectrum and, where possible,
uncertainties in baseline subtraction (note that for most spectra,
both authors independently computed these values).  Calibration
error is not included, since it is less than these other
uncertainties.  The peak $T_{mb}$ and FWHM were determined
by fitting Gaussians to the line profiles. 
For those positions
where no line was detected, 3$\sigma$ upper limits are given
for the peak T$_{mb}$ and $\int T_{mb}\,dV$, based on the
observed scatter in the spectra.  

\begin{table}
 \caption{Line Strengths at each Position}
 \label{symbols}
 \begin{tabular}{ccccc}
  Position  & Integration & 
$T_{mb}~(peak) $   & $\Delta V$ & $\int T_{mb} dV$ \\
 & (sec) & (K) & (FWHM) & (K km s$^{-1}$) \\ \\
 &  \twtt & & & \\ \\
 (0,0) & 3000 & 0.053 & 175 & 11.6 $\pm$ 0.95 \\
 ($-$7,+7) & 2400 & 0.055 & 380 & 18.7 $\pm$ 1.4 \\
 (+7,+7) & 2400 & 0.072 & 170 & 13.1 $\pm$ 1.2 \\
 (+7,$-$7) & 2400 & 0.030 & 120 & 5.6 $\pm$ 1.5 \\
 ($-$7,$-$7) & 2400 & 0.034 & 300 & 8.7 $\pm$ 1.4 \\ \\
 &  \thto & & & \\ \\
 (0,0) & 5520 & 0.017 & 140 & 1.24 $\pm$ 0.36 \\ \\
 &  \twto & & & \\ \\
 (0,0) & 3720 & 0.036 & 180 & 8.90 $\pm$ 0.90 \\
 ($-$10,+15) & 1800 & 0.046 & 30 & 2.70 $\pm$ 1.00 \\
 (+10,+15) & 2400 & 0.025 & 145 & 6.14 $\pm$ 1.06 \\
 (+10,$-$15) & 2400 & 0.028 & 135 & 4.13 $\pm$ 0.90 \\
 ($-$10,$-$15) & 2400 & 0.019 & 220 & 4.65 $\pm$ 1.03 \\
 ($-$20,0)   & 3600 & 0.020 & 95 & 2.71 $\pm$ 0.76 \\
 (+20,0)   & 3000 & 0.038 & 330 & 12.23 $\pm$ 1.00 \\
 (0,+30)   & 5400 & $<$ 0.025 & -- & $<$ 1.8 \\ 
 ($-$30,+15) & 5520 & 0.011 & 85 & 1.13 $\pm$ 0.41 \\
 (+30,+15) & 4800 & 0.031 & 85 & 2.85 $\pm$ 0.28 \\
 (+30,$-$15) & 5400 & $<$ 0.015 & -- & $<$ 1.8 \\ 
 ($-$30,$-$15) & 6690 & $<$ 0.015 & -- & $<$ 1.8 \\
 ($-$20,+30) & 6000 & $<$ 0.020 & -- & $<$ 2.4 \\
 (+20,+30) & 6000 & 0.018 & 65 & 1.20 $\pm$ 0.23 \\
 (+20,$-$30) & 5400 & $<$ 0.025 & -- & $<$ 3.0 \\
 ($-$40,0)   & 5400 & 0.013 & 50 & 1.39 $\pm$ 0.62 \\
 (+40,0)   & 3000 & $<$ 0.043 & -- & $<$ 5.3 \\  
 (+60,0)   & 1800 &  0.010 & 125 & 1.97 $\pm$ 0.61 \\ 
 \end{tabular}
 \medskip
\end{table}

\begin{figure*}
\epsfysize 4.0 truein
\hfil\rotate[r]{\epsffile
{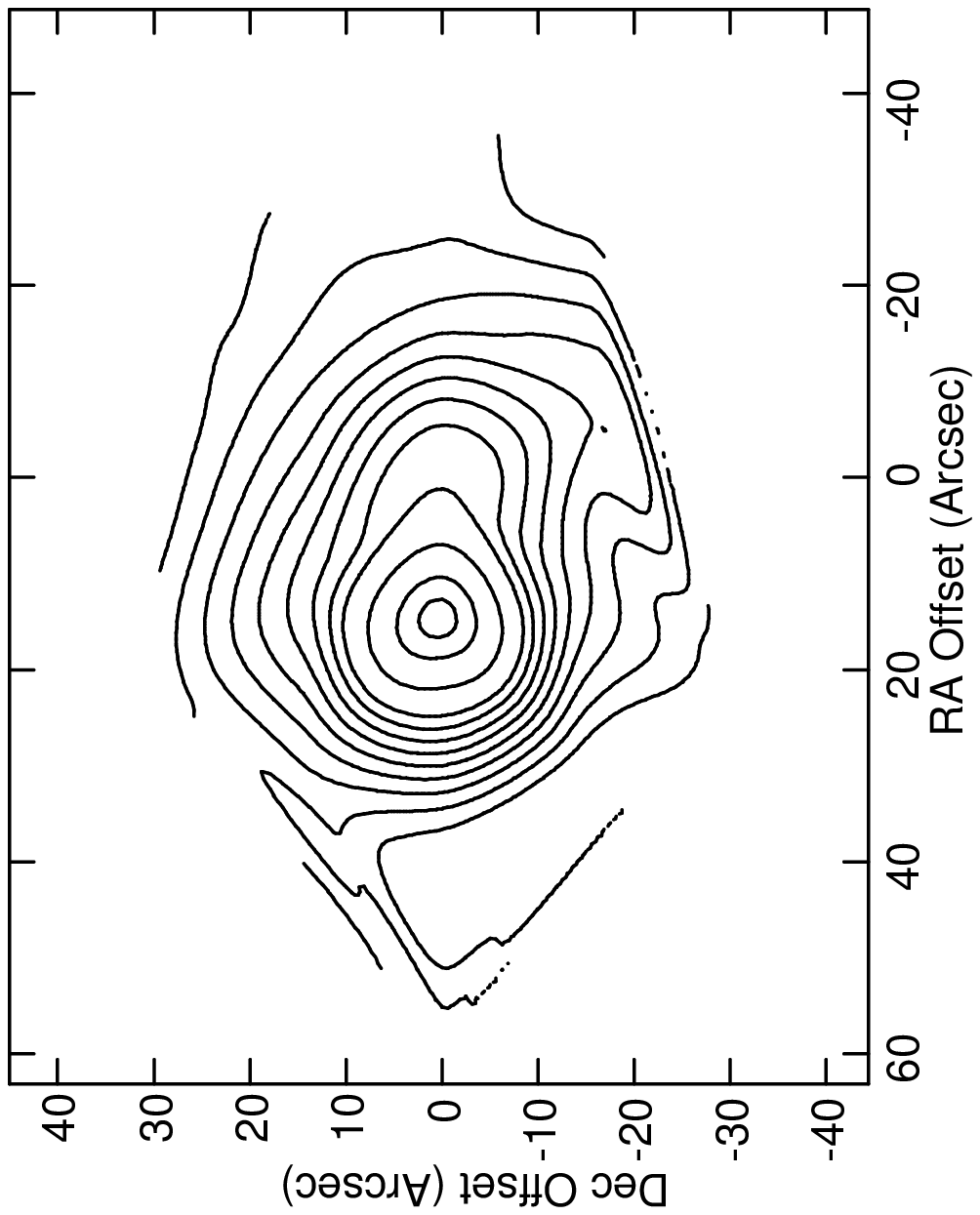}}\hfil
\epsfysize 4.0 truein
\hfil\rotate[r]{\epsffile
{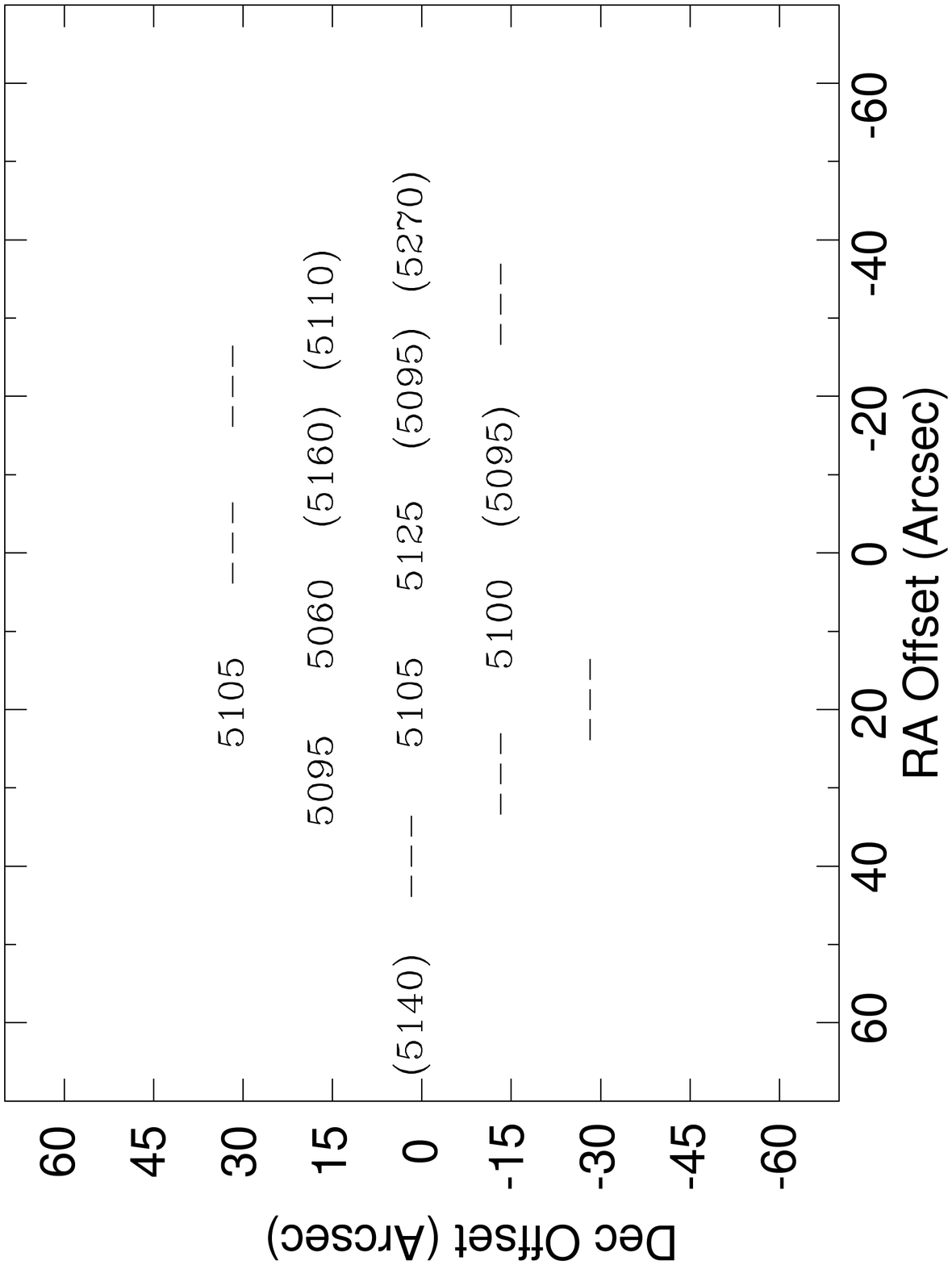}}\hfil
\caption{(a) Map of integrated intensities in
\twto. Contour levels are at 1, 2, 3, 4, 5, 6, 7, 8, 9,
11, 13, and 14 K km s$^{-1}$. 
 Structure around the extreme periphery is artificial due
to interpolation into regions containing no data. (b)
Crude Velocity Field in \twto.  The velocity has been
determined by finding the velocity centroids of each spectrum.
 Values enclosed in `(~~~)' have more uncertainty, and
`---' indicates that there is no line detected at that position.}
\label{fig4}
\end{figure*}

\subsection{The \twto Map and Velocity Field}

Our most extensive observations were obtained in the \twto line
and a map of integrated intensities is shown in Fig. 4a.  The
map was made by integrating each spectrum over velocity to
create a grid of $\int T_{mb} dV$ and then interpolating 
onto a finer grid using Langrange polynomials using
the Grenoble Image and Line Data
Analysis System (GILDAS) software package. 
Real detections are observed out to 36 arcsec (= 12 kpc)
from center.  

The main features in Figure 4a are an east-west elongation, a 
peak $\sim$ 15 arcsec
east of the galaxy center, and a smaller peak near the center
itself.  The east-west elongation has been observed in previous
maps, i.e. the single-dish data of Reuter et al. (1993) and
the interferometer data of Inoue et al. (1996).  The appropriate
comparison of our results is with 
previous single-dish \twto data, i.e. that
obtained by Reuter et al. at a resolution of 12$\sec$ (their Fig. 4).
They find three peaks,
one at (0,0),  one roughly coinciding with our strongest peak to the east,
and another $\sim$ 13\arcsec~ to the North-West.  After smoothing their
data to match our 21$\sec$ beam, the IRAM data shows a peak 
about 5$\sec$ to the west of centre.  Thus, there is some discrepancy 
between data sets in the sense that the JCMT data show a stronger peak
to the east of center whereas the IRAM data show a stronger peak to
the west of centre.  We cannot reconcile the data sets using different
baseline subtractions.  
Possible reasons for the discrepancy could include
the different
beam-size, coverage, and sampling between the two data sets.
Thus the general east-west nature of the emission
is confirmed but the relative intensities of the east-west peaks
requires further observations.
  
There is good overall agreement between the 
CO emission, the
low-velocity H$\alpha$ emission, 
the dust, and the hot X-ray
gas.   All share an East-West extension and have a similar spatial extent.
Like our \twto distribution (Fig. 4a), both
the X-ray emission (Bohringer et al. 1993)
as well as the low-velocity H$\alpha$ emission (Unger et al. 1990)
also display peaks
on the eastern side of the nucleus.
The radio map of
Pedlar et al. (1990) with $\sim$ 5\arcsec~ resolution, is oriented
at roughly 90 degrees to the CO and H$\alpha$ emission. 
As already mentioned, the radio emission seems to have displaced
the X-ray emitting gas (cf. Figure 1 of Bohringer et al.,
where the ROSAT HRI image is overlaid on the radio contours); we
return to this point in Section 5.2 and the Discussion.

Figure 4b shows a crude representation of the velocity
field in the \twto line,  where 
we have measured the velocity  
at each of the
positions in Figure 1, using the SPECX {\tt find-centroid} task.
Some of the spectra are low S/N, and we
have indicated this in Figure 4b by enclosing the velocity in 
parathenses [e.g. (5140)]; 
we have used `...' to indicate those 
positions for which no line was detected. 
Figure 4b  indicates that, not only is the \twto gas 
blueshifted at the (0,0) point, but all the gas over
the mapped area is 
blueshifted by 100--150 km s$^{-1}$ relative to the systemic velocity of
the galaxy.  (The same conclusion is reached if we instead plot the
peak velocities using gaussian fits to the spectra or whether we try
different baseline subtractions.)
This is in contrast to
Reuter et al. (1993) who found evidence for rotation from their
\twto velocity field (their Figure 5), but from their individual
\twto spectra (their Figure 2), it is not clear if this rotation
is significant.  Our \twto data show 
little evidence for rotation on the larger scales mapped. 
Since the central
\thto spectrum (Fig. 3) 
shows two peaks, consistent with the rotational 
components observed by Inoue et al. (1996), we find that 
any rotation, if present, appears
to be confined to a central ring.  We discuss this further in Section 6.2.

\section{The Molecular Gas: Line Ratios and Excitation Conditions}

\subsection{Formation of Line Ratios}

The IRAM \twto (beam size = 12$\sec$) and \twoz (beam size =
21$\sec$) $\int T_{mb} dV$ quantities of Reuter et al. (1993) have
been sampled in a 7$\sec$ grid in the central regions and an
11$\sec$ grid for the outer points.  Their calibration error
is estimated to be $\approx$ 10\% or lower for \twoz and $\approx$
10 - 20\% for \twto (Reuter, private communication).  We
interpolated these data using GILDAS in the same way as the JCMT data
(see \S3) and reproduced the maps
of Reuter et al. (their Fig. 3 and Fig. 4).  The data were then
smoothed, as required, to a 21 arcsec gaussian beam.

For the JCMT \twtt data (beam size = 14$\sec$), the sampling was
at 7 arcsec intervals
covering a 28$\sec$ region.  These data were also
interpolated and smoothed.
 For these data, it was also possible to form a line
ratio over a smaller 14$\sec$ beam, since the IRAM \twto 12$\sec$
resolution map could be smoothed to 14$\sec$.

Since there was some redundancy in the observations (i.e. \twto for
both IRAM and the JCMT), we could also check on the reliability of
the absolute calibration.  To do this, we smoothed the 
interpolated 12$\sec$ resolution IRAM \twto map to 21$\sec$ and directly
compared the emission within 21$\sec$ centered at (0,0) to the JCMT
\twto value at (0,0) in a 21$\sec$ beam.  The result is excellent
(possibly fortuitous) agreement to within 5\%. 

The resulting line ratios
for the (0,0) position are listed in Table 3, where we use the nomenclature,
$^{n}$CO(i-j) $\equiv$ $\int T_{mb}$[$^{n}$CO(i-j)]dV.
The final error bars in the ratio are of
order 15\% to 50\% and attempt to  take reasonable account of noise
in the spectra, differences introduced by subtracting different
baselines or choosing different integration limits, differences
introduced by interpolation/sampling, and differences (where
measured) between telescopes.  In the case of \thto, we also include
a measured variation between the ratio of integrated intensity
and ratio of peak intensity (Fig. 3).

\begin{table}
 \caption{CO Line Ratios
          }
 \label{symbols}
 \begin{tabular}{@{}lcccccc}
  Line Ratio &  Beam FWHM 
        & Value \\
 \twtt/\twto  &  14$\sec$  &   1.0 $\pm$ 0.15 \\
 \twtt/\twto  &  21$\sec$  &   1.25 $\pm$ 0.25 \\
\twto/\twoz  &  21$\sec$  &   0.74 $\pm$ 0.11 \\
 \twto/\thto  &  21$\sec$  &   6.0 $\pm$ 3.0 \\
 
 \end{tabular}
 \medskip
\end{table}

\subsection{Discussion of Line Ratios}

Our measured \twto/\twoz ratio of 0.74 $\pm$ 0.11 (Table 3) for
the cD galaxy, NGC~1275,
  falls within the range normally observed for spiral galaxies.
  For example, the
mean for optically
selected nearby spirals is \twto/\twoz = 0.89 $\pm$ 0.06, with
higher values ($\sim$ 1.1) measured for perturbed galaxies
(Braine \& Combes 1992; Braine et al. 1993).  Values of
0.93 $\pm$ 0.22 are found for IR bright galaxies (Aalto et al. 1995)
 and $>$ 1 for some starburst galaxies (Loiseau et al. 1990;
Braine et al. 1993).  Wilson, Walker, \& Thornley (1997) find values ranging between
0.5 to 1.07 for specific molecular cloud regions 
 in M33.  

The isotopic ratio of 
\twto/\thto = 6 $\pm$ 3 (based on our marginal detection of \thto)
is also typical of
giant molecular clouds (GMCs), for example, values
of 4 - 14 with a median of $\sim$ 5 - 6 are found 
for this ratio in the Galaxy
(Sanders et al. 1993) and 
Wilson, Walker, \& Thornley (1997)
find an average of 7 $\pm$ 1 for 7 molecular clouds 
in M33.  Higher ratios tend to be found for starburst galaxies
(e.g. 12 for M~82; Tilanus et al. 1991) and IR-bright and luminous
mergers ($>$ 20 for the CO(1-0) isotopic ratio; Aalto et al. 1995).
 
In contrast, our \twtt/\twto of 1.25 $\pm$ 0.25 (21$\sec$ beam) is
higher than normally observed in spirals.  Typically, the
\twtt/\twto ratio is lower than \twto/\twoz.  For example, Sanders
et al. (1993) find \twto/\twoz and \twtt/\twto ratios of $\sim$ 0.8 
(ranging from
$\sim$ 0.7 to 0.95) and $\sim$ 0.4 (ranging from $\sim$ 0.35 to
0.7), respectively, for Galactic GMCs within the solar circle.
Mean values for the nuclear region of IC~342
are 1.0 and 0.47 (Irwin \& Avery 1992).

Thus our line ratios for the giant cD galaxy, NGC~1275 
(Table 3) are not untypical of those observed
for normal spiral galaxies, with the possible exception of 
the \twtt/\twto ratio, which appears to be somewhat high.
Similarly high values of \twtt/\twto (i.e. 1.1 to 1.3)
have also been observed by Wall et al. (1993) in a small
sample of IR bright galaxies.  They require a 2-component model to
explain their data.  We explore this possibility in the next section.

\subsection{LVG Analysis and Cloud Properties}

For a line ratio analysis,
it is usually assumed that the observed main beam
brightness temperature, $T_{mb{_\nu}}(V)$, 
in a particular spectral line of frequency, $\nu$,
for a cloud moving at some velocity, $V$,
is just the radiation
temperature of the clouds, T$_{R_\nu}(V)$ (see Kutner \& Ulich 1981),
 diluted by a filling factor:
$$T_{mb{_\nu}}(V) \,=\, ff_\nu(V)\,*\,T_{R_\nu}(V) --- [1] $$ 
where the filling factor,
$ff_\nu(V)$, includes the area filling factor for clouds at velocity,
$V$, and a filling factor in velocity (due to dilution within 
channels) if required.  

In order to make use of the observed  (integrated) 
line ratios (Table 3),
it is necessary to assume that
the emission from each line
at velocity, $V$, is coming from
equivalent regions within the beam, i.e. 
the filling factors for each line at velocity, $V$, are equivalent.
This assumption also ensures that the ratios of
integrated line strength (which have higher S/N)
 are equal to the ratios
of peak $T_R(V)$ which is the modeled quantity.

We could check this assumption for the two cases
for which the original (unsmoothed) beam sizes were identical.  
The first is 
the JCMT \twto and IRAM \twoz data. 
A measurement of the peak T$_{mb}$\twto/T$_{mb}$\twoz ratio gives
a result which agrees with the integrated ratio (Table 2) to
within 5\%, well within the quoted errors.  The second is the JCMT
\thto and \twto data.
Since the \thto line may be more optically thin than
the \twto line, this line ratio is 
more likely than the others to suffer from differing filling factors.
This indeed appears to be the case, given the relative
 symmetry of the \thto line with respect to the
systemic velocity, in contrast to the observed asymmetry of
the \twto line (\S 3.2).  In this case, therefore, the error bar in
the \twto/\thto
line ratio takes into account the variation between the
ratio of peak T$_{mb}$ and the ratio of integrated line strengths.

Thus, within the uncertainties quoted and the
limitations of what can be checked, the assumption of equivalent
filling factors appears to be valid.

The line ratios of Table 3, applicable to a 21$\sec$ beam,
were considered in the context of
the Large Velocity Gradient (LVG) approximation
(Goldreich \& Kwan 1974; De Jong, Chu, \& Dalgarno 1975),
following Irwin \& Avery (1992) and using the collisional
rates of Flower \& Launay (1985) which are tabulated
from 10 K to 250 K. An 11-level CO molecule is assumed;
the excitation temperature of the highest level is 305 K.
Given a kinetic temperature,
$T_k$,
molecular hydrogen density, n$_{H_2}$, and fractional 
CO abundance, X, per unit
velocity gradient, (dV/dr), X (dV/dr)$^{-1}$, as inputs, the
radiation temperature, T$_R$,
optical depth, $\tau$, and excitation temperature,
T$_{exc}$, can be predicted for each line, as well as the line 
 ratios. The input quantities can then be adjusted until
the observed line ratios are reproduced.  

This method makes as few assumptions as possible about
the physical conditions in the gas (in comparison to, say, an
LTE model), an approach which is
particularly important for a cooling flow galaxy in which
 physical conditions may be atypical.  This approach does, however,
assume that the molecular clouds within the beam have
equivalent physical properties.
i.e. the beam-averaged line ratios provide us with
mean beam-averaged
physical properties of all clouds.
  We require only an
isotopic abundance ratio which we
take to be I = [\twco]/[\thco] = 50 $\pm$ 30
(Langer \&
Penzias 1990; also van Dishoeck \& Black 1988).  A velocity model
of the individual clouds is also required.  In practise,
the choice of velocity model is not critical, given the
uncertainties of the data;  here we assume
uniformly collapsing clouds (cf. Richardson 1985).

We searched through parameter space over a
range of
density from 10$^3$ to 10$^8$ cm$^{-3}$ and abundance 
per unit velocity gradient from 10$^{-10}$ to
10$^{-2}$ (km s$^{-1}$ pc$^{-1}$)$^{-1}$, both in steps of
10$^{n/2}$, n an integer, kinetic temperature from
10 K to 250 K in steps of 10 K, and for isotopic
abundance ratios of 20, 50, and 80.   The best-fit 
(from a $\chi^2$ test) one-component
results are listed in Table 4, column 2.  Since the \thto line
was only a marginal detection, we then repeated this process, excluding
this line.  These results are listed in Table 4, column 3.

From Table 4, we find similar results, regardless of the inclusion of
the isotopic line ratio, except for $\tau$.  
The results suggest the presence of cold
(20 - 30 K) clouds of low (10$^3$ cm$^{-3}$) density. 
A comparison
between the predicted line ratios of Table 4
(last 3 rows) and the observed
values of Table 3, however, shows that even the best-fit result
does not reproduce the observed \twtt/\twto and \twto/\twoz
ratios to within the quoted errors.
Again, if the isotope is excluded, the agreement is still poor.
  We cannot exclude the possibility that systematic
errors may be present, causing us to underestimate the uncertainties of
Table 3.  However, given the unusual environment of NGC~1275
(e.g. the presence of the radio core, and the possibilities of
the cooling flow and
past interaction), and considering the beam size (21$\sec$ = 7.14 kpc),  
we feel it is more likely that the clouds in the centre of this
galaxy cannot be described by a  one-component model.


\begin{table}
 \caption{Best-Fit One-Component LVG Results
          }
 \label{symbols}
 \begin{tabular}{@{}lcccccc}
  Parameter      & With \thto & Without \thto \\

T$_{kin}$ (K) &  30 & 20 \\
${n_H}_2$ (cm$^{-3}$) & 1 $\times$ 10$^{3}$  & 3 $\times$ 10$^3$ \\
X {(dV/dr)}$^{-1}$ (km s$^{-1}$ pc$^{-1}$)$^{-1}$ &  
1 $\times$ 10$^{-4}$ & 1 $\times$ 10$^{-4}$\\
I & 80 & --- \\

T$_R$[\twtt] (K) &  16.5 & 12.5 \\
T$_R$[\twto)] (K) &  20.6 & 14.6 \\
T$_R$[\thto] (K) &  2.7  & --- \\
T$_R$[\twoz] (K)  &  22.7 & 16.4 \\

$\tau$[\twtt]  & 42.3 & 148 \\
$\tau$[\twto] & 29.9 & 130 \\
$\tau$[\thto] &  1.5 & --- \\
$\tau$[\twoz] & 10.1 & 48.7 \\

T$_{exc}$[\twtt] (K) &  23.9 & 19.7 \\
T$_{exc}$[\twto] (K) & 26.0 & 19.8 \\
T$_{exc}$[\thto] (K) & 8.0 & --- \\
T$_{exc}$[\twoz] (K)  & 26.2 & 19.9 \\

\twtt/\twto &  0.80 & 0.86 \\
\twto/\twoz &  0.91 & 0.89 \\
\twto/\thto & 7.6 & --- \\

 \end{tabular}
 \medskip
\end{table}

A simple two-component model could be described in which
two types of clouds, A and B, are allowed, each with a unique value of
n(H$_2$),
X (dV/dr)$^{-1}$, T$_{kin}$ and filling factor.  Equation [1]
could then be replaced by:
$$T_{{mb}_\nu}\,=\,ff_A* T_{{R_A}_\nu} \,+\,ff_B*
T_{{R_B}_\nu} --- [4]$$
and the line ratios then computed as before.
We have indeed developed such an LVG model which 
can accommodate two components.
However, this two-component model now has 7, rather than 3, free
parameters and we no longer have sufficient data to solve for these
parameters.  We have, nevertheless, searched through parameter space
(where we have allowed the filling factor of component B compared to
A to range in steps of 10$^{n/4}$, n an integer from -25 to 0) to
see if there are regimes which can be excluded.  

With a two-component fit, it is possible to match the
observed line ratios quite well within error bars.  As expected, 
the results for the two-component model are generally unconstrained and
solutions are found over the entire range of density, temperature and
filling factor.  However, the one conclusion that can be reached 
from the two-component analysis is that
{\it it is not possible to find a solution for which the two cloud components
have the same temperature.}
Varying the density, filling factors, and/or 
abundance per unit velocity gradient between the two components
with a single temperature cannot reproduce
the observed line ratios.  (This conclusion remains, if the isotopic ratio
is excluded from the analysis.) 
Our best (but non-unique)
 two-component model which shows the {\it least} difference in
kinetic temperature has a cold (10 K) Component A at a density of
1.0 $\times$ 10$^{3}$ cm$^{-3}$ and a hot (170 K) Component B at a
density of 3.1 $\times$ 10$^4$ cm$^{-3}$ with the filling factor of
the hot component about half that of the cold component.
Thus the simplest interpretation of the results is
that temperature gradients occur in the 
molecular gas within the
central 21$\sec$ of NGC~1275.   This will be discussed
further in Section 6.


\section{Quantity and Pressure of the Molecular Gas}

\subsection{Total Molecular Gas Mass}

We have estimated the total molecular (H$_2$) gas in the usual way, using 
a Galactic conversion between CO intensity and M$_{H_2}$:

$$M_{H_2}~=~5.82 \times 10^6~I_{CO(1-0)}(\pi/4)d^2~~~,$$ 

\noindent where d is the telescope beam diameter in kpc
at the source distance
(here d=7.1 kpc for D=70 Mpc within our 21\arcsec~ beam).  
The numerical factor is taken from
Sanders, Solomon, \& Scoville (1984), and is applicable for the 
CO(1$-$0) transition.   We follow 
McNamara \& Jaffe (1994) and use the above equation for other
transitions by multiplying the numerical factor by the appropriate
line ratios taken from Table 3.  Our results can be summarized as
M$_{H_2}$~=~2.8 $\times$ 10$^9$ 
M$_\odot$ $\pm$ (8$-$10) $\times$ 10$^8$ M$_\odot$
for the \twoz, \twto, and \twtt transitions, where uncertainties 
in both the distance and the line ratios have been accounted for;
{\it no} uncertainty has been assumed in the Sanders, Solomon, \& Scoville CO(1$-$0)
numerical factor.  A similar result is found for the weaker 
\thto transition, but with a correspondingly larger uncertainty. 
The total mass over all emission mapped (shown by Fig. 4a, Table 2) is
M$_{H_2}$~=~1.6 $\times$ 10$^{10}$ M$_\odot$. 

We can compare our M$_{H_2}$ mass of $\sim$ 3 $\times$ 10$^9$ M$_\odot$
within our central 21$\sec$ beam
directly with that of Lazareff et al. (1989),
and Reuter et al. (1993) who observed NGC 1275 
with the 30m IRAM telescope in the \twoz transition with the same
beamsize, and found M$_{H_2}$~=~6 $\times$ 10$^9$ M$_\odot$ (all
comparisons are made assuming a distance of 70 Mpc
and adjusted to our conversion factor)
and  2.7 $\times$ 10$^9$ M$_\odot$, respectively.  
The latter authors find a total
mass of M$_{H_2}$~=~
8.6 $\times$ 10$^9$ M$_\odot$ over their total mapped region 
(a total area of $\sim$ 40$\sec$ $\times$ 30$\sec$ which is smaller
than that shown in Fig. 5).
Mirabel, Sanders, \& Kazes
(1989) obtained \twoz data with the 12m NRAO telescope, and 
found M$_{H_2}$~=~3.2 $\times$ 10$^9$ M$_\odot$ within their
55\arcsec~ beam.  
Finally, Inoue
et al. (1996) find M$_{H_2}$ of 8.6 $\times$ 10$^9$ and 
4.3 $\times$ 10$^{10}$ M$_\odot$ 
within diameters of 9\arcsec and 65\arcsec
respectively.   Thus, our M$_{H_2}$ mass is comparable with 
most previous single dish measurements, though the interferometer
data give higher results. 


We must stress that the above
determinations of the molecular gas mass have relied
upon a Galactic conversion between M$_{H_2}$ and CO integrated
intensity.  Conditions in cluster cooling flows are likely to be very 
different from those of Galactic molecular
clouds, and the conversion factor may depend on such factors as
metallicity and the intensity of the UV radiation field (e.g.
Wilson 1995; Arimoto, Sofue, \& Tsujimoto 1996). 
Further observational and theoretical work is needed to determine the
conversion factor over the full range of environments encountered (early and
late type galaxies, starbursts, cooling flows, etc.). 

\subsection{Gas Pressure}

Unfortunately we cannot compute the pressure of the molecular gas with
certainty because we have only rough constraints on the gas
temperature and density from our LVG analysis (\S 4.5).  However, we can
use the values from our best two-component solution
(Section 4.3) to provide a sample solution. 
Taking T$_{kin}$=170 K and n$_{H_2}$=
3 $\times$ 10$^4$ cm$^{-3}$ for the hot, dense Component B 
(the pressure from Component A can be ignored), we find a molecular
gas pressure P $\simeq$ 7 $\times$ 10$^{-10}$ erg cm$^{-3}$.  Since
Component B's density is reasonably well constrained
and its tabulated temperature is a lower
limit (\S 4.3), this result should represent a rough lower limit, in
the context of the two-component LVG model.
Bohringer et al. (1993) quote a pressure for the hot X-ray
emitting gas in the central region of 
Perseus of 2.5 $\times$ 10$^{-10}$
erg cm$^{-3}$, while C. Peres (private communication) finds P $\sim$ 2.5 
$\times$ 10$^{-9}$ erg cm$^{-3}$
at a radius of 8\arcsec, both from ROSAT HRI data (and both
values corrected to H$_0$=75).   The latter number is
the best to compare with our value, and the agreement is reasonable
given the uncertainties in both our LVG analysis and the X-ray
determinations near the cluster center.  We can also compare with
the pressure found for the optical [SII] emission lines by Heckman
et al. (1989): P$_{opt}$=7 $\times$ 10$^{-10}$ erg cm$^{-3}$ between
2$-$5 kpc (6$-$15\arcsec).  Thus, there is good agreement between the
gas pressures as determined from X-ray, sub-mm, and optical data.

Bohringer et al. (1993) overlaid their ROSAT HRI image of Perseus with
the 332 MHz radio data of Pedlar et al. (1990), and found that the
X-ray gas and radio plasma are anti-correlated.
In particular, the
outer radio lobes are coincident with minima in the X-ray surface
brightness, and Bohringer et al. speculated that the radio plasma
has displaced the thermal gas.   Similar results are found for other
clusters with lobe-dominated radio sources, 
including Abell 1795, 2029, 2597 and 4059 (see Sarazin
1997 for references), Abell 2634 (Schindler \& Prieto 1997),
and Abell 2199 (Owen \& Eilek 1997).

\section{Discussion}

\subsection{Displacement of the Molecular Gas by the Radio Plasma?}

We indicated in Section 5.2 that the radio lobes and thermal gas have
similar pressures, and that the radio lobes and X-ray gas avoid each
other.  Following Bohringer et al. (1993),
we speculate that the radio lobes are thus displacing the
hot X-ray gas, the optical line-emitting gas, {\it and} the molecular
gas.  If the molecular gas on the side of the galaxy nearest to us is
being pushed outwards, this could explain the blueshift of 100$-$150 
km s$^{-1}$
we observe in the \twto and \twtt emission.  We now show that this idea can
also work energetically. 

 Roughly speaking, the kinetic energy needed to
displace the $\sim$ 1.6 $\times$ 10$^{10}$ M$_\odot$ 
($\times$ 1.4 to account for heavier elements) of molecular gas at a
velocity of 150 km s$^{-1}$ is $\sim$ 
5 $\times$ 10$^{57}$ ergs; compare this with the
total (minimum) energy in the radio halo (Component A)
 of 3 $\times$ 10$^{57}$ ergs
determined by Pedlar et al. (1990). The latter radio value applies to
a region 38$\sec$ $\times$ 15$\sec$ in size and would actually be
higher over a region which is comparable to the molecular gas 
component [Fig. 4(a)] since more radio emission is present on
larger scales.  A simple linear scaling for area gives a modified radio
energy of $\sim$ 16 $\times$ 10$^{57}$ ergs, or a factor of $\sim$ 3
higher than the molecular gas kinetic energy in the same region.  Thus
(depending on the efficiency of energy transfer and given
the uncertainties) there appears to
be sufficient energy in the radio components to account for the
velocity displacement of the molecular gas.

Note that it is unlikely that the blueshifted emission represents
gas on the far side of the nucleus which is infalling as a result of
the cooling flow because velocities as high as 150 km s$^{-1}$ are
not expected for quasistatically contracting cooling flow gas.

\subsection{What's Special About NGC 1275?}

\bigskip

Why is NGC 1275/Perseus the only cooling flow cluster with detected
CO emission?
The best upper limits on M$_{H_2}$
are already a few times 
10$^9$ M$_\odot$ (e.g.  McNamara \& Jaffe 1994; 
Braine \& Dupraz 1994; O'Dea et al. 1994), and there are at
least 14 clusters which would have been detected in CO if they
contained quantities of  
molecular gas similar to that found in NGC 1275/Perseus. 
As we discussed in the Introduction, there have been
detections of excess X-ray absorption in many clusters (White 
et al. 1991; Allen \& Fabian 1997).  While the
physical properties
of this absorbing material 
are not at all known (see discussion in Sarazin 1997), 
it has been argued that much of it might consist of very cold
($\sim$ 3K) and dense molecular clouds.  
However, NGC 1275
happens to have one of the strongest radio cores
of any cooling flow central galaxy.  This radio core may heat the
molecular gas to detectable temperatures in NGC 1275/Perseus, while
the gas remains cold in most other clusters.  As we have discussed
already, the radio plasma is certainly affecting the thermal gas on
larger scales.

There are other possibilities.  Braine et al. (1995) point out that
``The FIR and CO emission from NGC 1275 correspond exactly to what is 
found in gas-rich spirals.  Rather than a massive cooling flow,
the gas may come from accretion of one or more gas-rich galaxies."
There is evidence for a past merger or interaction in NGC 1275 from
the shells and young globular clusters seen in HST images  
(note, for instance, that 
NGC 1275 is probably suffering an interaction with the high-velocity
system seen in absorption).  In this case, the molecular gas would
be much warmer and hence detectable.  As well, the gas could be heated
up during the merger event itself.  
Mergers appear to be rare in rich clusters at
the present epoch, given their large
velocity dispersions (e.g. Merritt 1985), 
and this is consistent with the non-detections
of CO in other cooling flow clusters.  A merger of a dust and gas
rich spiral
could also be the
source of the thermal emission from dust observed in NGC 1275
(e.g. Lester et al. 1995).

\section{Summary}

NGC~1275 is a remarkable galaxy with properties not seen anywhere else. 
With its central position in the Perseus cooling flow, its young
blue globular clusters, the evidence for a past merger, the foreground
infalling system, and its molecular gas component, it it so far unique
amongst all galaxies, not just those of its class.  Here we have
focussed on the molecular gas distribution in NGC~1275, since this is
the {\it only} cooling flow galaxy (so far) which displays a cold
molecular component.  Using the JCMT, we have 
mapped the \twto distribution
of this galaxy at a resolution of 21$\sec$, detecting gas out to
radii of 36$\sec$ (12 kpc).  Within the limitations of our spatial resolution
and coverage, we find an approximate east-west orientation to the
emission, consistent with that found by previous authors in H$\alpha$
emission, X-ray emission, and the dust component.  Evidence for the
inner molecular rotating disk is also seen, consistent with NMA
interferometric data.  However, on broader scales we see no evidence
for rotation.

The CO line profiles are consistently bluewards asymmetric, both at
the central position in both \twtt and \twto
and over the entire region mapped in \twto, where the S/N is
sufficient to measure this quantity. We interpret this blueshift
as being due to kinetic energy imparted by the radio jet/outflow
in the advancing direction on the near side of the nucleus.
With an estimate of total H$_2$
mass (assuming Galactic conversion factors) of 1.6 $\times$ 10$^{10}$
M$_\odot$, and velocity offset of 150 km s$^{-1}$, we find that
$\sim$ 5 $\times$ 10$^{57}$ ergs is required
to impart this kinetic energy. There appears to be sufficient energy
in the radio source to account for this, given the uncertainties.
This result is consistent with Bohringer et al. (1993) who found
that the X-ray gas has been displaced by the radio jet.  No 
corresponding
redshifted component is seen, possibly because the (receding) radio
jet is less active.

Together with additional JCMT \thto data as well as
\twto and \twoz IRAM data from Reuter et al.
(1993), we have formed three line ratios applicable to a central
21$\sec$ beam.  Two of the line ratios appear to be typical of
cold Galactic clouds, but the third [\twtt/\twto] is suggestive
of gas under more ``active", warmer conditions.  This is confirmed
by an LVG radiative transfer analysis which indicates that the
observed line ratios cannot be reproduced by a single component, i.e.
a single set of kinetic temperature, molecular hydrogen density,
and abundance per unit velocity gradient.  We have therefore 
developed a simple two-component model to describe the data.
In the two-component analysis, there are insufficient line ratios
to fully contrain the set of cloud parameters; however, the results
indicate that the observed line ratios can be reproduced if
the two molecular components are at different temperatures.
  It may be that a temperature
gradient exists in the central region possibly due to the radio core.

Finally, while we note that the characteristics of the
molecular gas in NGC~1275 have been 
readily related to the nuclear radio activity,
these observations do not allow us to rule out
a possible connection with the cooling flow, itself.
\bigskip
\bigskip
\bigskip

\section*{acknowledgements}

\noindent{The authors wish to thank Dr. Lorne Avery and
Jessica Arlett for providing the initial LVG code.  We are grateful
to H.-P. Reuter and his collaborators
for sharing their IRAM data, and to
Clovis Peres for showing us his ROSAT analysis
before publication.  We appreciate very much the assistance we
received from our Support Astronomer Chris Purton, and Bill Dent,
Per Friberg, and Tim Jenness who carried out remote observing for us
in December 1995.   Thanks to Henry Matthews for answering many 
questions relating to the 
calibration of our data. 
Thanks finally to Rachael Padman for all of
her work in developing the SPECX package, and for helping
us to use it.  We thank the anonymous referee for 
a thorough reading of the paper, and suggestions for 
improvement.  This work
has been supported by Natural Sciences and Engineering Research
Council of Canada Grant \# OGP0184201 (for JI).}

\end{document}